\documentclass{ismdproc}

\begin{document}

\title{Power-law ensembles: fluctuations of volume or temperature ?}

\author{{\slshape Grzegorz Wilk$^1$,
                  Zbigniew W\l odarczyk$^2$,
                  Wojciech Wolak$^3$
                  \footnote{Speaker: Grzegorz Wilk$^1$ (email: wilk@fuw.edu.pl)}}\\[1ex]
$^1$The Andrzej So{\l}tan Institute for Nuclear Studies,
Ho\.{z}a 69, 00681, Warsaw, Poland\\
$^2$Institute of Physics,Jan Kochanowski University,
\'Swi\c{e}tokrzyska 15, 25-406 Kielce, Poland\\
$^3$Kielce University of Technology, Tysi\c{a}clecia Pa\'nstwa
Polskiego 7, 25-314 Kielce Poland}

\contribID{xy}  
\confID{yz}
\acronym{ISMD2010}
\doi            

\maketitle

\begin{abstract}
  We present two issues here: (i) that in situation when total energy
  is kept constant recently proposed fluctuations of volume ensemble
  is equivalent to the approach using Tsallis statistics with fluctuating
  temperature and (ii) that the later
  (in which fluctuations are described by the nonextensivity parameter
  $q$) leads to the observed experimentally sum rule connecting fluctuations
  of different physical observables.
  \end{abstract}

\section{Introduction}
\label{section:I}

Statistical modelling represents a standard tool widely used to
analyze multiparticle production processes \cite{MG_rev}. However,
this approach does not account for the possible intrinsic
nonstatistical fluctuations in the hadronizing system which
usually result in a characteristic power-like behavior of single
particle spectra or in the broadening of the corresponding
multiplicity distributions (and which can signal a possible phase
transition(s) \cite{PhTr}). To include such features one should
base this modelling on the so called Tsallis statistics
\cite{T,WW_epja,BPU_epja} (represented by Tsallis distribution)
which accounts for such situations by introducing, in addition to
the temperature $T$, one new parameter, $q > 1$, directly
connected to fluctuations \cite{WW,BJ} (for $q \rightarrow 1$ one
recovers the usual Boltzmann-Gibbs distribution):
\begin{equation}
\exp_q \left(-\frac{E}{T}\right) = \left[1 -
(1-q)\frac{E}{T}\right]^{\frac{1}{1-q}}\, \stackrel{q \rightarrow
1}{\Longrightarrow}\, \exp \left(-\frac{E}{T}\right),\qquad
q-1=\frac{Var(1/T)}{\langle 1/T\rangle^2} .\label{eq:example}
\end{equation}
 The most recent applications of this approach come from PHENIX
Collaboration at RHIC \cite{PHENIX} and from CMS Collaboration at
LHC \cite{LHC_CMS}. One must admit at this point that this
approach is subjected to a rather hot debate of whether it is
consistent with the equilibrium thermodynamics or it is only a
handy way to phenomenologically description of some intrinsic
fluctuations in the system under consideration \cite{debate}.
However, as was recently demonstrated on general grounds in
\cite{M}, fluctuation phenomena can be incorporated into
traditional presentation of thermodynamic and Tsallis distribution
\cite{T} belongs to the class of general admissible distributions
which satisfy thermodynamical consistency conditions and which are
therefore a natural extension of the usual Boltzman-Gibbs
canonical distribution. Actually, what was shown in \cite{WW} was
that starting from some simple diffusion picture of temperature
equalization in the nonhomogeneous heat bath (in which local $T$
fluctuates from point to point around some equilibrium
temperature, $T_0$) one gets evolution of $T$ in the form of
Langevin stochastic equation and distribution of $1/T$, $f(1/T)$,
as solution of the corresponding Fokker-Planck equation. It turns
out that $f(1/T)$ has form of gamma distribution,
\begin{eqnarray}
f(1/T) = \frac{1}{\Gamma\left(\frac{1}{q - 1}\right)}\frac{T_0}{q
- 1}\left(\frac{1}{q - 1}\frac{T_0}{T}\right)^{\frac{2 - q}{q -
1}}\cdot \exp\left( - \frac{1}{q - 1}\frac{T_0}{T}\right).
\label{eq:gamma}
\end{eqnarray}
Convoluting $\exp (- E/T)$ with such $f(1/T)$ one gets immediately
Tsallis distribution, $\exp_q(-E/T)$ from Eq. (\ref{eq:example})
\cite{WW}. Parameter $q$, i.e., according to Eq.
(\ref{eq:example}) also the temperature fluctuation pattern, is
therefore fully given by the parameters describing this basic
diffusion process (cf., \cite{WW} for details, this was recently
generalized to account for the possibility of transferring energy
from/to heat bath, which appears to be important for AA
applications \cite{WW_epja,WWprc} and for cosmic ray physics
\cite{WWcosmic}; we shall not discuss this issue here). This
approach has now been successfully applied in many circumstances,
see \cite{WW_epja,WWprc,PHENIX,LHC_CMS} and references therein.

\section{Fluctuations of $V$ or $T$?}
\label{section:II}

It must be stressed at this point that the form of $f(1/T)$ as
given by Eq. (\ref{eq:gamma}) is not assumed but has been derived
from the underlying physical process. We shall now compare this
approach with that proposed in \cite{Vfluct} in which the volume
$V$ was assumed to fluctuate in the scale invariant way following
the observed KNO scaling behavior of the multiplicity
distributions, $P(N)$ \cite{KNO}. We shall demonstrate here that
when total energy is kept constant, as was assumed in
\cite{Vfluct}, both approaches are equivalent. Let us first notice
that for constant total energy, $E=const$, both the volume $V$ and
temperature $T$ are related via $E \sim VT^4$, what means that
\begin{equation}
T = \langle T\rangle\left( \frac{\langle V \rangle}{V}
\right)^{\frac{1}{4}} = \frac{\langle T\rangle}{y}\quad {\rm
where}\quad y = \left( \frac{V}{\langle
V\rangle}\right)^{\frac{1}{4}}. \label{eq:TV}
\end{equation}
Following now \cite{Vfluct} the mean multiplicity in the
microcanonical ensemble (MCE), $\bar{N}$,  can be written as
\begin{equation}
\bar{N} = \langle N\rangle\cdot \frac{V}{\langle V\rangle}\left(
\frac{T}{\langle T\rangle}\right)^3\ =\ \langle N\rangle\ y,
\label{eq:barNV}
\end{equation}
what means that $\bar{N}$ fluctuates in the same way $y$. It is
then natural to assume that $y$ follows the pattern of
fluctuations of $\bar{N}$, i.e., KNO limit of the NBD distribution
observed in data fitting \cite{KNO}, which is given by gamma
function. The power-like form of single particle spectra then
follow immediately, all apparently without invoking any reference
to Tsallis statistics. Notice, however, that because of
(\ref{eq:TV}), $1/T$ will fluctuate according to the same gamma
distribution. So we get $T$ fluctuations with the same functional
form but now without the physical background behind the Eq.
(\ref{eq:gamma}) mentioned above. However, we can proceed in
reverse order and obtain from our $T$ fluctuations introduced in
Section \ref{section:I} fluctuations of $V$ introduced in
\cite{Vfluct}. In this sense both approaches are equivalent with
the former being based on some physical processes and the second
on apparently ad hoc assumption\footnote{However, after all, this
assumption can have some phenomenological foundation, not
mentioned in \cite{Vfluct}, which deserves further scrutiny.
Namely, one observes experimentally a variation of the emitting
radius (evaluated from the Bose-Einstein correlation analysis)
with the charged multiplicity of the event, see, for example,
\cite{R_fluct}. An increase of about $10$ \% of the radius when
the multiplicity increases from $10$ to $40$ charged hadrons in
the final state was reported. Unfortunately, the quality of data
does not allow us to precisely determine the power index of the
volume dependence. It is also remarkable that both the energy
density, $\rho_E=E/V$, and particle density, $\rho_N = N/V$,
decrease for large multiplicity events. For $N/\langle N\rangle
\sim y$ one observes $\rho_E/\langle \rho_E\rangle \sim y^{-4}$
and $\rho_N/\langle \rho_N\rangle \sim y^{-3}$. All these deserves
further consideration and should be checked in future LHC
experiments, especially in ALICE, which is dedicated for heavy ion
collision.}.

We close this section with short reminder that temperature
fluctuations discussed in Section \ref{section:I} result in
automatic broadening of the corresponding multiplicity
distributions, $P(N)$, from the poissonian form for exponential
distributions to the negative binomial (NB) form for Tsallis
distributions \cite{fluct}. It is known that whenever we have $N$
independently produced secondaries with energies $\{
E_{i=1,\dots,N}\}$ taken from the  exponential distribution in Eq.
(\ref{eq:example}) and whenever $\sum^N_{i=0} E_i \leq E \leq
\sum^{N+1}_{i=0} E_i$, then the corresponding multiplicity
distribution is poissonian,
\begin{equation}
P(N) = \frac{\left( \bar{N}\right)^N}{N!} \exp\left ( -
\bar{N}\right) \quad {\rm where}\quad \bar{N} =\frac{E}{\lambda}.
\label{eq:Poisson}
\end{equation}
What was shown in \cite{fluct} is that whenever in some process
$N$ particles with energies  $\{ E_{i=1,\dots,N}\}$ are
distributed according to the joint $N$-particle Tsallis
distribution,
\begin{equation}
h\left(\{ E_{i=1,\dots,N}\} \right)\! =\! C_N\left[ 1-
(1-q)\frac{\sum^N_{i=1} E_i }{\lambda} \right]^{\frac{1}{1-q}+1-N}
\label{eq:NTsallis}
\end{equation}
(for which the corresponding one particle Tsallis distribution
function in Eq. (\ref{eq:example}), is marginal distribution),
then, under the same condition as above, the corresponding
multiplicity distribution is the NB distribution,
\begin{equation}
P(N)\, =\, \frac{\Gamma(N+k)}{\Gamma(N+1)\Gamma(k)}\frac{\left(
\frac{\langle N\rangle}{k}\right)^N}{\left( 1 + \frac{\langle
N\rangle}{k}\right)^{(N+k)}};\quad {\rm where}\quad
k=\frac{1}{q-1}.\label{eq:NBD}
\end{equation}
Notice that in the limiting cases of $q\rightarrow 1$ one has
$k\rightarrow \infty$ and (\ref{eq:NBD}) becomes a poissonian
distribution (\ref{eq:Poisson}), whereas for $q\rightarrow 2$ on
has $k\rightarrow 1$ and (\ref{eq:NBD}) becomes a geometrical
distribution. It is easy to show that for large values of $N$ and
$\langle N\rangle$ one obtains from Eq. (\ref{eq:NBD}) its scaling
form,
\begin{equation}
\langle N\rangle P(N) \cong  \psi\left( z=\frac{N}{\langle
N\rangle} \right) = \frac{k^k}{\Gamma(k)} z^{k-1}\exp( - kz),
\label{eq:scalingform}
\end{equation}
in which one recognizes a particular expression of
Koba-Nielsen-Olesen (KNO) scaling \cite{KNO} and which, as
discussed before, has been assumed to describe also the volume
fluctuations in \cite{Vfluct}. Here it results from the
temperature fluctuations described by the parameter $q$ discussed
in Section \ref{section:I} with well defined physical
meaning\footnote{It is worth to mention at this point that, as
shown in \cite{RWW}, fluctuations of $\bar{N}$ in the poissonian
distribution (\ref{eq:Poisson}) taken in the form of
$\psi(\bar{N}/<N>)$, Eq. (\ref{eq:scalingform}), lead to the NB
distribution (\ref{eq:NBD}).}.

\section{Composition of different fluctuations}
\label{section:III}

Description of fluctuations phenomena by means of the parameter
$q$ using Tsallis statistics allows for better understanding of
interrelations between different fluctuations. From our experience
with $p{\bar p}$ collisions \cite{compq} we know that one can
obtain very good description of the whole range of $p_T$ ($\propto
\exp_q\left( -p_T/T \right)$ with $(T_T$ [GeV]$;q_T)=
(0.134;1.095)$, $(0.135;1.105)$ and $(0.14;1.11)$ for energies (in
GeV) $200$, $540$ and $900$, respectively. These values should be
compared with the corresponding values of $(T=T_L;q=q_L)$ obtained
when fitting rapidity distributions ($\propto \exp_q\left( -\mu_T
\cosh y/T\right)$) at the same energies: $(11.74;1.2)$,
$(20.39;1.26)$ and $(30.79;1.29)$. It was noticed there that $q_L
-1$ has the same energy behavior as $1/k$ in the NB distribution
fitting the multiplicity distributions at corresponding energies
($q_L - 1 = -0.104 + 0.058 \ln \sqrt{s}$). It means that
fluctuations of total energy are in this case driven mainly by
fluctuations in the longitudinal phase space. Explanations
proposed in \cite{compq} was following. Noticing that $q-1 =
\sigma^2(T)/T^2$ (i.e., it is given by fluctuations of total
temperature $T$) and assuming that $\sigma^2(T) = \sigma^2(T_L) +
\sigma^2(T_T)$, one can estimate that resulting values of $q$
should not be too different from
\begin{equation}
q\, =\, \frac{q_L\, T_L^2\, +\, q_T\, T^2_T}{T^2}\, -\,
        \frac{T^2_L\, +\, T^2_T}{T^2}\, +\, 1 \quad \stackrel{T_L \gg
        T_T}{\Longrightarrow}\quad \sim q_L.
\label{eq:qqq}
\end{equation}

\begin{figure}[h]
\includegraphics[width=0.5\textwidth]{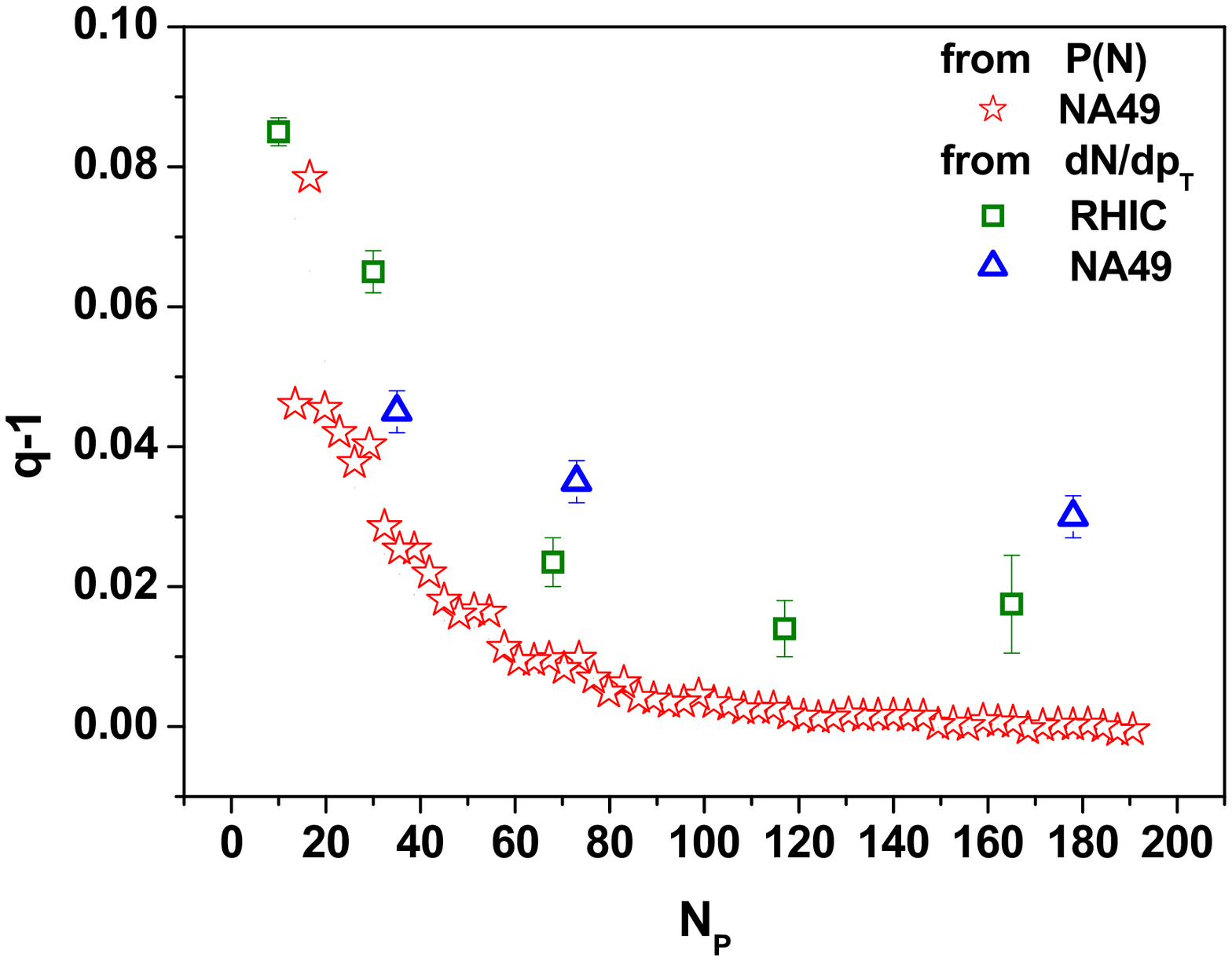}
\includegraphics[width=0.5\textwidth]{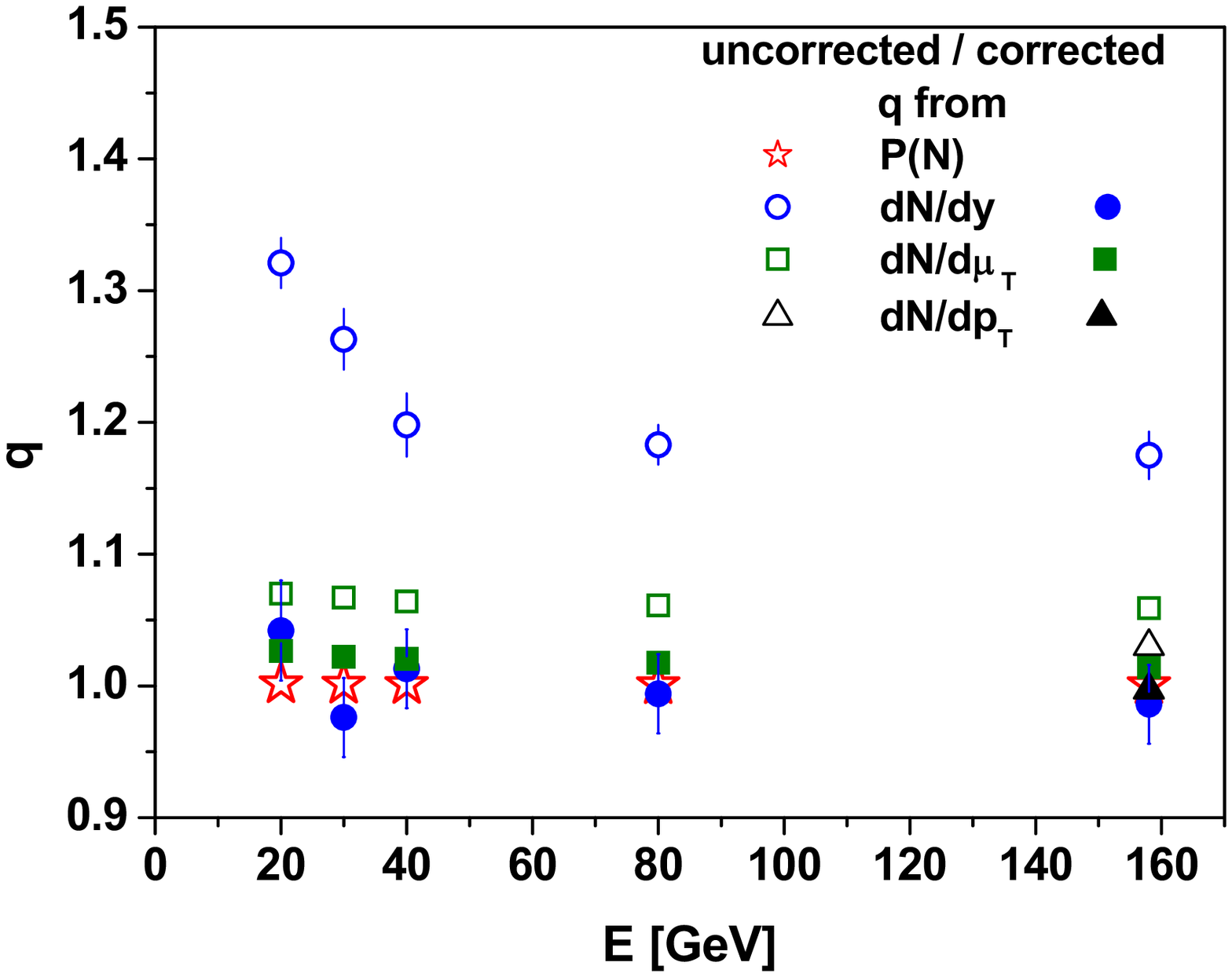}
\caption{ Left panel: $q$ from $P(N)$ are our results obtained in
\cite{WWprc} from $Var(N)/\langle N\rangle$, $q$ from RHIC are
taken from compilation \cite{qcompilation} and are based on data
for $dN/dp_T$ from  \cite{RHIC} analyzed there, whereas $q$ from
NA49 are obtained by us for this presentation using data on
$dN/dp_T$ from the first work in \cite{NA49}. Right panel: all
results were obtained for the sake of this presentation using
distributions provided by \cite{NA49}, i.e., respectively,
$dN/d\mu_T$, $dN/dy$ and $dN/dp_T$. The errors are similar to
those presented as example for $q$ obtained from $dN/dy$. Open
symbols correspond to uncorrected values of $q$, full symbols to
values corrected by means of the procedure proposed in the
text.}\label{Fig1}
\end{figure}

It turns out that situation is completely reversed in the case of
nuclear collisions, which we shall discuss now\footnote{We use for
this purpose NA49 data on $Pb+Pb$ collisions \cite{NA49} because,
at the moment, only this experiment measures at the same time (at
least for the most central collisions) multiplicity distributions,
$P(N)$, and distributions in rapidity $y$, transverse momenta,
$p_T$, and transverse masses, $\mu_T=\sqrt{m^2 + p^2_T}$, which is
crucial for our further considerations here.}, cf. Fig.
\ref{Fig1}. Left panel shows $q$ obtained from different sources
as function of centrality represented by number of participants,
$N_P$. The one obtained from $P(N)$ follows
\begin{equation}
q - 1 = \frac{1}{aN_P}\left( 1 - \frac{N_P}{A} \right)
\label{eq:qP(N)}
\end{equation}
behavior ($a = 0.98$) \cite{WWprc}. Whereas for small centralities
it approaches situation encountered in $p\bar{p}$ collisions
(where it was practically equal to $q=q_L$ obtained from rapidity
distributions as mentioned above), the more central is event the
smaller is $q -1$, i.e., the nearer to poissonian is the
corresponding $P(N)$. Notice that both, $q_L$ and $q_T$ (obtained
from $p_T$ distributions are now greater than $q$ and have
(approximately) visible similar dependence on $N_P$, however now
$q_L < q_T$, again opposite to what was seen in $p\bar{p}$
\cite{compq} (for comparison $q_T$ obtained by \cite{qcompilation}
using RHIC data $Au+Au$ collisions at $200$ GeV \cite{RHIC} are
shown here as well). Right panel shows the same quantities but now
as function of energy for the most central $Pb+Pb$ collisions
\cite{NA49}. In both cases we take from \cite{NA49} distributions
of rapidity, $dN/dy$, and in $\mu_T$, $dN/d\mu_T$, and from them
deduces the corresponding $q$.

The natural question is, what causes such different behavior of
parameter $q$ in this case. The answer we propose is the
following. When extracting values of parameter $q$ from the
rapidity distributions tacit assumption was made that $\mu_T$ in
$E = \mu_T\cosh y$ remains constant (i.e., it does not fluctuate).
What would happen if this assumption was false? Notice that in the
$\exp_q\left( -E/T\right) = exp_q\left[ -
\left(\mu_T/T\right)\cosh y\right] = \exp_q( - z\cosh y)$. It
means that fits to rapidity distributions provide us, in fact
fluctuations not so much of partition temperature $T$ but rather
of the variable $z = \mu_T/T$. This in turn can be written
approximately as:
\begin{equation}
Var(z) \simeq \frac{1}{\langle T\rangle^2}Var\left( \mu_T\right) +
\frac{\langle \mu_T \rangle^2}{\langle T\rangle^2} \cdot
\frac{Var(T)}{\langle T\rangle^2}. \label{eq:varz}
\end{equation}
Because $\langle z\rangle \simeq \langle \mu_T\rangle/\langle
T\rangle$ and $Var(1/T)/\langle 1/T\rangle^2 \simeq Var(T)/\langle
T\rangle^2$ and because $Var(z)/\langle z\rangle^2 = Var\left(
\mu_T\right)/\langle \mu_T \rangle^2 + Var(T)/\langle T\rangle^2$
one can write that
\begin{equation}
q - 1 \stackrel{def}{=} \frac{Var(T)}{\langle T\rangle^2} =
\frac{Var(z)}{\langle z\rangle^2} - \frac{Var\left(
\mu_T\right)}{\langle \mu_T\rangle^2} . \label{eq:sumrule}
\end{equation}

This sum rule is our main result and its action is presented in
the right panel of Fig. \ref{Fig1}. It connects total $q$, which
can be obtained from the analysis of the NB form of the measured
multiplicity distributions, P(N), with $q_L -1 = Var(z)/\langle
z\rangle^2$, obtained from fitting rapidity distributions and
$Var\left(\mu_T\right)/\langle \mu_T\rangle^2$ obtained from data
on transverse mass distributions. When extracting $q$ from
distributions of $dN/d\mu_T$ we proceed in analogously way with
$z$ being in this case equal to $z=\cosh y/T$.

\section{Summary}
\label{section:IV}

To summarize: we have demonstrated that for constant total energy
fluctuations of $T$ introduced by us some time ago
\cite{WW,WW_epja} are equivalent to fluctuations of $V$ proposed
recently \cite{Vfluct} and that, at the moment, the former have
advantage of being backed by a plausible physical arguments
\cite{WW}. Moreover, due to relation (\ref{eq:TV}) valid for
constant total energy, the inverse temperature $1/T$ and $V^{1/4}$
fluctuate in the same way, according to gamma distribution and
such fluctuations lead to the Tsallis form of the respective
distributions for energy spectra.

However, the main results presented here is the sum rule formula,
Eq. (\ref{eq:sumrule}), connecting $q$ obtained from analysis of
different distributions which are obtained {\it in the same
experiment}. This allows us to understand why in $AA$ collisions
fluctuations observed in multiplicity distributions are much
smaller than the corresponding ones seen in the rapidity
distribution or in distribution of transverse momenta (i.e., why
the corresponding $q$ parameters evaluated from distributions of
different observables are different). This issue should be checked
further when completely sets of data would become available from
the experiments at LHC (especially from ALICE).

\section*{Acknowledgements}
Partial support (GW) of the Ministry of Science and Higher Education under contract DPN/N97\\
/CERN/2009 is acknowledged.

\begin{footnotesize}

\end{footnotesize}

\end{document}